\documentclass[fleqn,usenatbib]{mnras}

\usepackage[T1]{fontenc}
\usepackage{ae,aecompl}
\usepackage{graphicx}
\usepackage{epstopdf}
\usepackage{amsmath}
\usepackage{amssymb}
\usepackage{enumerate}

\title[Energetic Scaling of Black Hole Jets]{Extending the ``Energetic Scaling of Relativistic Jets From Black Hole Systems" to Include $\gamma$-ray-loud X-ray Binaries}
\author[G.P. Lamb, S. Kobayashi, and E. Pian]{Gavin P Lamb$^{1}$, Shiho Kobayashi$^{1}$, Elena Pian$^{2, }$$^{3}$
\\
$^{1}$Astrophysics Research Institute, LJMU, IC2, Liverpool Science Park, 146 Brownlow Hill, Liverpool L3 5RF, 
UK\\
$^{2}$INAF IASF Bologna, Via P. Gobetti 101, 40129 Bologna, Italy\\
$^{3}$Scuola Normale Superiore de Pisa, Piazza dei Cavalieri 7, I-56126 Pisa, Italy}
\date{Accepted XXX. Received YYY; in original form ZZZ}
\pubyear{2017}

\begin{document}
\label{firstpage}
\pagerange{\pageref{firstpage}--\pageref{lastpage}}
\maketitle

\begin{abstract}
We show that the jet power $P_j$ and geometrically corrected $\gamma$-ray luminosity $L_\gamma$ for the X-ray binaries (XRBs) Cygnus X-1, Cygnus X-3, and V404 Cygni, and $\gamma$-ray upper limits for GRS 1915+105 and GX339-4, follow the universal scaling for the energetics of relativistic jets from black hole (BH) systems found by \cite{2012Sci...338.1445N} for blazars and GRBs.
The observed peak $\gamma$-ray luminosity for XRBs is geometrically corrected;
and the minimum jet power is estimated from the peak flux density of radio flares and the flare rise time.
The $L_\gamma-P_j$ correlation holds across $\sim 17$ orders of magnitude.
The correlation suggests a jet origin for the high energy emission from X-ray binaries, and indicates a common mechanism or efficiency for the high energy emission 0.1-100 GeV from all relativistic BH systems.
\end{abstract}

\section{Introduction}

Astrophysical jets are observed on many different scales from proto-stars and X-ray binaries (XRBs) within our Galaxy, to radio-galaxies, blazars, and $\gamma$-ray bursts (GRBs) at cosmological distances.
Relativistic jets from black hole (BH) systems have a broad range of luminosities and dynamics:
XRBs with a BH component have bolometric luminosities that can reach $\sim 10^{39}$ erg s$^{-1}$ \citep[e.g.][]{1999ARA&A..37..409M,2004ARA&A..42..317F}, with Lorentz factors constrained by observations of the jet and counter-jet to a few $\Gamma\leq 5-10$;
blazars, the on-axis analogue to the kilo-parsec jet structures of radio-galaxies \citep{1995PASP..107..803U}, have luminosities of $\sim 10^{48}$ erg s$^{-1}$, and Lorentz factors $\Gamma\leq 40-50$ \citep[e.g.][]{2006AIPC..856....1M, 2009AJ....138.1874L,2013ApJ...773..147J,2016MNRAS.461..297S};
GRBs have energy outputs $\sim10^{52}$ erg s$^{-1}$, where achromatic temporal breaks in the afterglow indicate a jet structure \citep[e.g.][]{1999ApJ...519L..17S}, and Lorentz factors $\ga 100$ can be inferred from the highly variable non-thermal emission \citep[e.g.][]{2002ARA&A..40..137M,2004RvMP...76.1143P}.

Several attempts have been made to unify the different scales of BH engines.
The relativistic jets or ouflows from BH systems are thought to have a common mechanism.
The appearance of superluminal features in a jet following a dip in X-ray emission has been observed for  both XRBs and the radio-galaxy 3C120, where the X-ray dip is associated with accretion \citep{2002Natur.417..625M}.
A fundamental plane connecting BH mass, radio, and X-ray luminosity was found for active galactic nuclei (AGN) and XRBs by \cite*{2003MNRAS.345.1057M}.
A scaling relation for the radio flux, from the core of AGN and XRBs, with BH mass $M$ (or accretion rate), where most accretion scenarios produce the relation $F_\nu \propto M^{17/12-s/3}$ and $s$ is the spectral index where $s=0$ for flat spectrum sources and $s\sim 0.75$ for optically thin emission, demonstrates that the radio-loudness of jets scales with BH mass, where the mass can range over nine orders of magnitude \citep{2003MNRAS.343L..59H}.
Similarly, scaling laws have been found to unify low-power accreting BH over many decades in mass \citep{2004A&A...414..895F}.
The emission models for jets from a supermassive BH have also been successfully applied to an XRB e.g. GRS 1915+105 \citep{2004A&A...415L..35T}.

Comparisons between the jets from different mass BH systems led to \cite{2012ApJ...757...56Y} using the nature of episodic jets from AGN and XRBs to explain the erratic light-curves of GRBs.
A correlation between blazar jets and GRBs was demonstrated by \cite{2012Sci...338.1445N};
by considering the power of GRB and blazar jets $P_j$, and the collimation corrected $\gamma$-ray luminosity $L_\gamma$, the relation $P_j \propto L_\gamma^{0.98}$ was found.
Blazars and GRBs occupy the low and high ends of the correlation respectively.
This result implies that the efficiency of the $\gamma$-ray producing mechanism within these jets is consistent over 10 orders of magnitude in jet power.

There have been several attempts to find a unifying scheme or scaling relation between BH systems where accretion and ejection is at work.
Many results have been obtained that separately relate AGN and GRBs, or AGN and XRBs.
An attempt to relate all three classes was made recently by \cite{2017MNRAS.470.1101W}, they used the X-ray and radio luminosities from GRBs and inferred a BH mass to show that the fundamental plane of BH activity \citep{2003MNRAS.345.1057M,2004A&A...414..895F} holds for all jetted BH systems.
Also, by considering the bolometric luminosity from jets, \cite{2014ApJ...780L..14M} demonstrated that BH XRBs and low-luminosity AGN fit on the $L_\gamma-P_j$ relation for GRBs and AGN.
If the $L_\gamma-P_j$ relation is truly universal then the on-axis and collimation corrected $\gamma$-ray luminosity and power for the jets from XRBs should fit the same relation as for blazars and GRBs.
A fit to this relation could indicate a ubiquitous emission mechanism for all relativistic BH jets and allow for constraints on the high energy emission models for XRBs, AGN, and GRBs.

The XRBs Cygnus X-1 \citep{2013ApJ...775...98B,2016A&A...596A..55Z,2016arXiv160705059Z}, Cygnus X-3 \citep{2013ApJ...775...98B,2012MNRAS.421.2947C}, and V404 Cygni \citep{2016MNRAS.462L.111L} have been detected at {\it Fermi} LAT $\gamma$-ray energies.
A further two sources have {\it Fermi} LAT upper limits;
GRS 1915+105 and GX339-4 \citep{2013ApJ...775...98B}.
All of these objects have evidence for a BH component \citep{2016ApJS..222...15T,2016A&A...587A..61C}:
Cygnus X-1 (Cyg X-1), has a BH confirmed by dynamical modelling \citep{2011ApJ...742...84O};
Cygnus X-3 (Cyg X-3), has a radio and X-ray correlation which follows that found in BH X-ray binaries \citep{2008MNRAS.388.1001S};
V404 Cygni (V404 Cyg) has a BH confirmed by the mass function \citep{1994MNRAS.271L...5C};
GRS 1915+105, the BH is established using a dynamical mass estimate \citep{2014ApJ...796....2R};
GX339-4, the K-correction and model confirm the BH \citep*{2008MNRAS.385.2205M}.
Including these XRB on the $L_\gamma-P_j$ universal scaling found by \cite{2012Sci...338.1445N}, we make the first attempt, using $\gamma$-ray luminosities, at comparing the energetics for three classes of accreting BH systems.
The comparison is extended to $\sim17$ decades in both $\gamma$-ray luminosity and jet power.

In $\S$\ref{S2} the XRB parameters are discussed.
$\S$\ref{S3} outlines the method for correcting the $\gamma$-ray luminosity and inferred jet power for the inclination and collimation.
The results are presented in $\S$\ref{S4}.
The discussion and conclusion are in $\S$\ref{S5} and $\S$\ref{S6}.

\section{XRB Parameters}
\label{S2}

The inclusion of XRBs on the $L_\gamma-P_j$ relation requires estimates for the $\gamma$-ray luminosity from the relativistic jets, for an on-axis observer, and estimates for the jet power.
Unlike blazars and GRBs, the jets from XRBs are not guaranteed to be oriented along the line-of-sight.
Any detected emission from an off-axis jet will have to be corrected for the relativistic Doppler effect;
this requires knowledge of the system inclination and bulk Lorentz factor $\Gamma$.
Additionally, any high-energy emission from the jet will be collimated within an angle $1/\Gamma$, which is typically greater than the jet half-opening angle for XRBs.
To estimate the jet power we assume equipartion of energy between the particles and magnetic field, and use the optically thin emission during radio flares to find the minimum power. 
The necessary parameters are:
the detected $\gamma$-ray photon flux $N$;
the system distance $D$;
the jet inclination $i$;
the jet bulk Lorentz factor $\Gamma$;
and radio flare peak flux density $S_\nu$, observed frequency $\nu$, and rise time $\Delta t$.

Radio emission and flares from XRBs are attributed to relativistic jets.
Accretion, seen at X-ray energies, and ejection, seen at radio, are strongly correlated \citep[e.g.][etc.]{1998A&A...330L...9M,1998MNRAS.300..573F,2000A&A...359..251C,2001MNRAS.322...31F,2003A&A...400.1007C,2003ApJ...595.1032R, 2008ApJ...675.1449R,2013MNRAS.428.2500C}.
The peak flux and rise time of radio flares can be used to constrain the power of a jet.
Emission at $\gamma$-ray energies from XRBs has been associated with radio flaring and variability \citep{2012MNRAS.421.2947C}.
Detection of $\gamma$-rays during periods of intense radio flaring suggests the origin of the high energy emission is a jet \citep{2013ApJ...775...98B}.
The simultaneous detection of the 511 keV annihilation line and higher energy $\gamma$-rays from V404 Cyg within hours of a giant radio flare indicates a jet as the origin of the $\gamma$-ray emission \citep{2016MNRAS.462L.111L}. 

XRBs are not persistent $\gamma$-ray sources at detection sensitivity, although see \cite{2013ApJ...775...98B} where Cyg X-3 was detected above the background without a flare. 
Generally XRBs have only been observed at these high energies during flares; 
therefore we use the detected peak {\it Fermi} LAT $\gamma$-ray photon flux for each source and determine an observed isotropic equivalent $\gamma$-ray luminosity $L_{\gamma,{\rm obs,iso}}$ from the {\it Fermi} LAT photon spectral index\footnote{High energy photon spectral index is regularly represented using $\Gamma$; to avoid confusion with the outflow bulk Lorentz factor ($\Gamma$) we use $\alpha$ throughout} $\alpha$ at energies $>100$ MeV.
Detections are in the 0.1-10 GeV range for Cyg X-1 and Cyg X-3 \citep{2013ApJ...775...98B, 2009Sci...326.1512F}, and 0.1-100 GeV for V404 Cyg \citep{2016MNRAS.462L.111L}.
Upper limits for the $\gamma$-ray photon flux from GRS 1915+105 and GX339-4, in the energy range 0.1-10 GeV, are used to estimate the maximum $L_{\gamma,{\rm obs,iso}}$ for these objects \citep{2013ApJ...775...98B}.
The detected peak photon flux and spectral index $\alpha$, for Cyg X-1, Cyg X-3 and V404 Cyg, and the $\gamma$-ray photon flux upper limits for GRS 1915+105 and GX339-4 are shown in Table \ref{table1}.

The photon spectral index is defined as $N_E \propto E^{-\alpha}$, where $N_E$ is in units ph s$^{-1}$ cm$^{-2}$ erg$^{-1}$ and $E$ is the photon energy.
The $\gamma$-ray luminosity is then,
\begin{equation}
L_{\gamma,{\rm obs,iso}} \sim 1.9\times 10^{35} N_{-6} D_{\rm kpc}^2 \frac{(\alpha-1)}{(\alpha-2)}\frac{(E_{\rm low}^{2-\alpha}-E_{\rm high}^{2-\alpha})}{(E_{\rm low}^{1-\alpha}-E_{\rm high}^{1-\alpha})}~~~{\rm erg~s^{-1}},
\label{E1}
\end{equation}
where $E_{\rm low}$ and $E_{\rm high}$ are the detection band limits in GeV, $N_{-6}=N/(10^{-6} {\rm~ph~s^{-1}~cm^{-2}})$ and $N$ is the detected photon flux, and $D_{\rm kpc}$ is the distance in kpc.

The observed proper motion of radio jet components can be used to put constraints on the value of $\Gamma$.
The proper motion is defined as $\mu=c\beta\sin i/[D(1\pm\beta\cos i)]$ radians s$^{-1}$, where $\beta=(1-\Gamma^{-2})^{1/2}$.
An approaching component $\mu_a$ has $1-\beta\cos i$ and a receding component $\mu_r$ has $1+\beta\cos i$.
Using resolved $\mu_a$ and $\mu_r$, a value for $\beta\cos i$ can be found, where $\beta \cos i=(\mu_a-\mu_r)/(\mu_a+\mu_r)$ \citep{1999ARA&A..37..409M}.
Values of $\beta\cos i$ for various XRBs are listed by \cite*{2006MNRAS.367.1432M} (MFN06 from here).
For a system with a known inclination, the observable quantity $\beta\cos i$ can be used to determine the bulk Lorentz factor $\Gamma$.
Where the inclination is unknown, the Lorentz factor can be determined using the approaching and receding proper motions and the distance to the system.
From the product of the proper motions $\mu_a\mu_r$,
\begin{equation}
\Gamma = \left[1-x^2-\mu_a\mu_r\frac{D^2(1-x^2)}{c^2}\right]^{-1/2},
\label{E2}
\end{equation}
where $x$ is the observed value $\beta\cos i$, the proper motions $\mu_a$ and $\mu_r$ are in radians s$^{-1}$, $D$ is the distance in cm, and $c$ is the speed of light in cm s$^{-1}$.

If the proper motions of either component are poorly constrained then a limit on $\Gamma$ can be found by considering the observed jet opening angle $\phi$.
The angle $\phi$ is an upper-limit found by measuring the angle between the jet central axis and a tangential line from the edge of a radio component to the system core.
The jet components are assumed to be spherical plasmoids that expand uniformly with a co-moving velocity $\beta_{\rm exp}$.
If we assume maximum co-moving expansion velocity of $c$, then the jets bulk Lorentz factor is, $\Gamma \ga [1+\tan^{-2}\phi\sin^{-2}i]^{1/2}$.
Where the co-moving expansion velocity is less than the maximum, $\Gamma \ga [1+\beta_{\rm exp}^2/(\tan^2\phi\sin^2i)]^{1/2}$.
This assumes no jet confinement.

The inclination $i$ of the system to the line of sight is well constrained for Cyg X-1, V404 Cyg, and GRS 1915+105 \citep[respectively]{2011ApJ...742...84O,2017ApJ...834...90H, 2014ApJ...796....2R}.
Cyg X-3 and GX339-4 have unknown system inclinations.
For Cyg X-3; \cite{2010MNRAS.404L..55D} showed that the jet orientation within the system is constrained to be between $20^\circ \la \theta_j \la 80^\circ$, and the system line-of-sight inclination is $i=30^\circ$.
\cite{2013A&A...550A..48V} used an inclination of $i=30^\circ$ in their models.
Using the $\beta\cos i$ values in MFN06 and the distance to the system, the bulk Lorentz factor $\Gamma$ of the jet can be constrained.
Given $\beta\cos i=0.5$, $\mu_a\mu_r\sim 7.4\times10^{-26}$ rads s$^{-1}$, and $D=7$ kpc, the bulk Lorentz factor is $\Gamma=1.18$ and the line of sight inclination to the jet-axis is $i \simeq 20^\circ$.
For GX339-4; MFN06 measured $\beta\cos i \geq 0.16$ and derived a $\Gamma\geq4.9$ from the jet opening angle;
a lower limit of $\Gamma \geq 2.3$ is used by \cite*{2004MNRAS.355.1105F}.
Using these values for $\Gamma$, the inclination of the system can be determined from $\beta\cos i = 0.16$;
for $\Gamma=4.9$ the inclination is $i=80^\circ.6$;
for $\Gamma=2.3$ the inclination is $i=79^\circ.8$.
In all cases we assume that the inclination angle is the same as the line-of-sight angle to the jet axis and that there is no significant precession.
Values for the inclination are listed in Table \ref{table1}.

XRB jets typically have $\Gamma<5$.
For Cyg X-1 a Lorentz factor $\Gamma=1.25$ is used by \cite{2015A&A...584A..95P} for modelling the lepto-hadronic broadband emission,
whilst from the jet opening angle and $\beta\cos i$ from MFN06, there is a minimum value of $\Gamma=3.3$.
We show results for both values.
For Cyg X-3 the Lorentz factor must be $\Gamma\leq 2$ (MFN06);
we derived the value $\Gamma=1.18$ (equation \ref{E2}).
We show results for $\Gamma=2$ and $\Gamma=1.18$.
For V404 Cyg we assume a Lorentz factor $\Gamma=2.3$ \citep{2016ApJ...823...35T}.
For GRS 1915+105, from the inclination $i=60^\circ$ and $\beta\cos i=0.41$ we derive a $\Gamma=1.75$.
For GX339-4, we use the value $\Gamma=4.9$ from the jet opening angle. 

We assume that $\gamma$-ray emission, and radio flares are from the jet with negligible contribution from the accretion disk or star.
The peak radio flare flux density $S_\nu$, the rise time $\Delta t$, frequency $\nu$, and distance $D$ are shown with references in Table \ref{table1}.

\section{Method}
\label{S3}

We use the {\it Fermi} LAT measured $\gamma$-ray photon flux for three XRBs: Cygnus X-1 \citep{2013ApJ...775...98B}, Cygnus X-3 \citep{2009Sci...326.1512F}, and V404 Cygni \citep{2016MNRAS.462L.111L}.
These are currently the only $\gamma$-ray detected XRBs.
{\it Fermi} LAT upper limits exist for GRS 1915+105 and GX339-4 \citep{2013ApJ...775...98B};
the upper limits are used for these objects.
Cynus X-3 and V404 Cygni have also been detected at $>100 $ MeV by AGILE \citep{2009Natur.462..620T, 2017arXiv170310085P}.
The high energy emission is associated with jet activity.

Emission from a relativistic jet is beamed in the direction of the jet bulk motion;
we assume a point like emission region on the jet axis for all high energy photons.
The $\gamma$-ray luminosity is corrected for the inclination of the jet to the line of sight.
The Lorentz invariant quantity $I_\nu/\nu^3$ \citep{1986rpa..book.....R}, where $I_\nu$ is the specific intensity and $\nu$ the frequency, can be used to determine the specific luminosity from a relativistic source where the observer is outside the relativistic beaming angle.
As $\nu=\delta\nu'$, where $\delta=[\Gamma(1-\beta\cos i)]^{-1}$ is the relativistic Doppler factor, $\Gamma$ the bulk Lorentz factor, $i$ the inclination, and primed quantities are in the co-moving frame, then $I_\nu=I'_{\nu'}(\nu/\nu')^3=I'_{\nu'}\delta^{3}$.
The observed luminosity is then $L_\nu=4\pi I'_{\nu'}\delta^3$.
For an on-axis observer the Doppler factor becomes $\delta=[\Gamma(1-\beta)]^{-1}$;
the observed luminosity is then $a^3$ times the on-axis luminosity \citep{2002ApJ...570L..61G}, where $a$ is the correction for an on-axis observer to an off-axis observer;
the factor $a= (1-\beta)/(1-\beta\cos{i})$.

The $\gamma$-ray luminosity for an on-axis observer has detection band limits a factor $a^{-1}$ times the off-axis detection limits;
a correction to the on-axis Doppler boosted emission should be made to ensure the detection band is consistent.
All the $\gamma$-ray detections have a single power-law spectral fit with a $\nu F_\nu$ index $2-\alpha$, and no information of a spectral peak or behaviour at lower energies.
A peak for the $\gamma$-ray component should exist at a few GeV \citep[e.g.][]{2014MNRAS.442.3243Z} for an on-axis observer;
we therefore assume a flat spectrum for the correction.
The on-axis isotropic equivalent $\gamma$-ray luminosity is then $L_{\gamma,{\rm iso}}=a^{-3} ~L_{\gamma,{\rm obs,iso}}$, where $L_{\gamma,{\rm obs,iso}}$ is the observed isotropic equivalent $\gamma$-ray luminosity, equation \ref{E1}.
The collimation-corrected luminosity is $L_\gamma=f_b ~L_{\gamma,{\rm iso}}$, where $f_b$ is the collimation factor for the jet.
The collimation-correction is $f_b=1-\cos{(1/\Gamma)}$.
The intrinsic, on-axis $\gamma$-ray luminosity is then $L_\gamma=f_b~a^{-3}~L_{\gamma,{\rm obs,iso}}$.

Bright radio flares from plasmoids that travel along the relativistic jet structures can be used to estimate the minimum power of the jet.
Although $\gamma$-ray emission is often correlated with radio flaring, the site of the emission within the jet is distinct.
Radio flares are contained by the plasmoids and equipartition of the energy within these structures can be assumed.
The jet power is estimated by assuming equipartition of energy between the synchrotron emitting particles and the magnetic field strength $B$  \citep{1956Phys.Rev...103...264,1959ApJ...129...849,1994hea2.book.....L,2006csxs.book.....L}.
The energy density in the particles, given a random magnetic field, is $e \propto B^{-3/2}$, and the energy density in the magnetic field is $u \propto B^2$.
The total energy is $E_{\rm total}=V(e+u)$, where $V$ is the volume of the emitting region;
as the dominant component is unknown i.e. large $B$ and small $e$, or small $B$ and large $e$, then a minimum energy can be found at the point where ${\rm d}E_{\rm total}/{\rm d}B = 0$.
The particle number density assumes a power-law distribution of ultra-relativistic electrons $n_e \propto E^{-p}$;
the contribution from relativistic protons is included by the factor $\eta=1+\epsilon_p/\epsilon_e$, where $\epsilon_p$ is the energy in protons and $\epsilon_e$ is the energy in electrons.
The energy in the particles is $E = C(p,\nu) \eta L_\nu B^{-3/2}$, where $L_{\nu}$ is the co-moving specific luminosity and $C(p,\nu)$ is a constant that depends on the particle index $p$, the frequency $\nu$ of the specific luminosity, and the upper and lower synchrotron frequency limits for the particle distribution.

For a distribution of particles with a power-law index $p>2$, the low energy particles dominate.
By assuming that $\nu=\nu_{\rm min}$, the minimum synchrotron frequency, a simple estimate for the energy in the system can be made\footnote{This assumes no large flux of low energy relativistic particles with a different energy spectrum}.
We assume a particle distribution, in all cases, of $p=2.5$ \citep{2011ApJ...726...75S};
the observed flux density $S_\nu$ would have a spectral index of $0.75$, where $S_\nu\propto \nu^{-0.75}$.
The volume of the emitting system is assumed to be spherical, where the size can be inferred from the light crossing time indicated by radio flare rise time $\Delta t$;
the volume is then $V=4\pi(\Delta t c)^3/3$.
The jet-power $P_j=E_{\rm total}/\Delta t$ can then be estimated by considering the Doppler corrected observed flux density;
for an optically thin source the Doppler correction to the flux density is $\delta^{3+(p-1)/2}$ \citep{1979ApJ...232...34B}.

The jet power $P_j$ is a Lorentz invarient quantity, therefore the observed flux density, time, and frequency must be co-moving quantities.
The flux dependence is $S'_{\nu'} = \delta^{-(3+(p-1)/2)} S_\nu$, the time is $\Delta t' = \delta \Delta t$, and frequency $\nu' = \delta^{-1} \nu$.
The jet power is then,
\begin{equation}
P_j\sim 3.5\times 10^{33} ~\eta^{4/7} ~\Delta {t'}^{2/7} ~{\nu'}_{{\rm GHz}}^{2/7} ~{S'}_{{\nu'}{\rm, mJy}}^{4/7} ~D_{{\rm kpc}}^{8/7}~~~{\rm erg ~s}^{-1},
\label{PJ}
\end{equation}
where, $\Delta t'$ is in seconds, $\nu'$ is in GHz, $S'_{\nu'}$ is in mJy, and $D$ is in kpc.
We assume equal energy in protons and electrons, $\epsilon_p/\epsilon_e=1$.

Uncertainties on the derived values are estimated by propagating the uncertainty on the distance, the inclination, the $\gamma$-ray flux, and the bulk Lorentz factor.
The uncertainty on $\Gamma$ is assumed to be $d\Gamma=0.1$ for Cyg X-1, Cyg X-3, GRS 1915+105, and GX339-4 where the estimate for $\Gamma$ is from observed proper motions, and $d\Gamma=0.5$ for V404 Cyg where $\Gamma$ is found from a model jet velocity.
The choice of uncertainty for $\Gamma$ reflects the estimation method and a conservative value for the minimum precision.
The error on the final parameters is dominated by the uncertainty in the $\gamma$-ray flux and is only very weakly dependent on the choice of $d\Gamma$.

\section{Results}
\label{S4}

\begin{figure*}
\includegraphics[scale=0.8]{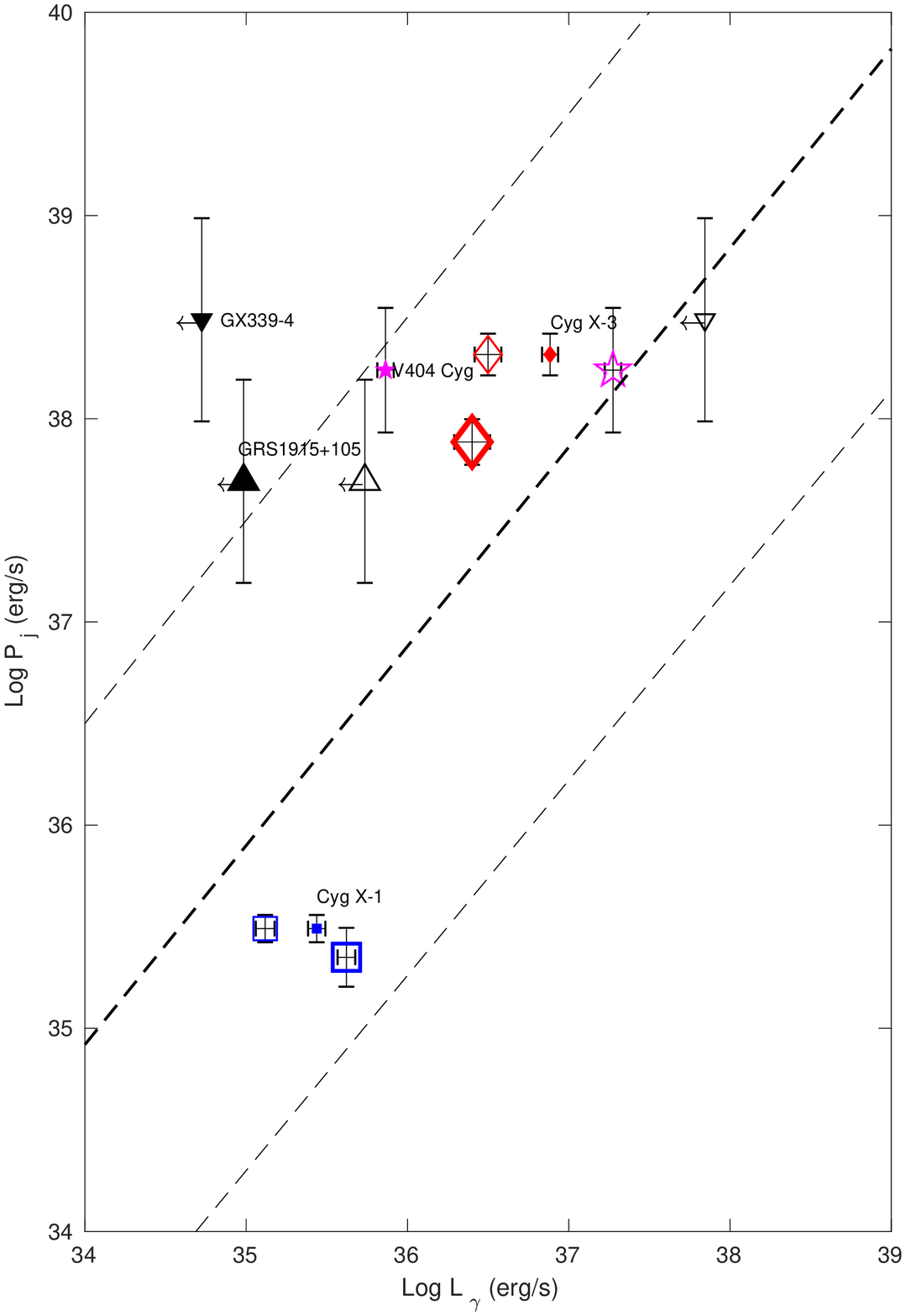}
\centering
\caption{XRB $L_\gamma-P_{j}$ diagram: Cyg X-1 blue squares; Cyg X-3 red diamonds; V404 Cyg pink stars; GRS 1915+105 black upward-pointing triangle; GX339-4 black downward-pointing triangle. Filled markers represent observed values with no $L_\gamma$ correction for collimation and Doppler-factor - where two values of $\Gamma$ are listed, the lowest value is used in determining the jet power. Unfilled markers represent the collimation- and Doppler-corrected values - small markers represent the lower $\Gamma$ value, large marker represents the larger $\Gamma$ value. Black solid lines indicate the uncertainties calculated for each value. $L_\gamma$ for GRS1915+105 and GX339-4 are upper limits. The dashed line is the $L_\gamma-P_{j}$ relation found by Nemmen et al. 2012 for Blazars and GRBs; the dotted lines are their uncertainties.}
\label{fig1}
\end{figure*}

Figure \ref{fig1} shows the $L_\gamma - P_{j}$ relation for the sample of XRBs.
The observed luminosities (filled markers) and collimation/Doppler-corrected values (unfilled markers) are both shown.
Values for Cyg X-1 are blue squares;
Cyg X-3 are red diamonds;
and V404 Cyg, are pink stars.
For Cyg X-1; the small unfilled marker is the estimate based on $\Gamma=1.25$, the large marker is $\Gamma=3.3$.
For Cyg X-3; the small unfilled marker is $\Gamma=1.18$, and the large $\Gamma=2$.
For V404 Cyg, there is only one estimate for the bulk Lorentz factor used.
GRS 1915+105 is an upward pointing black triangle.
GX339-4 is a downward pointing black triangle.
Errorbars are those derived from the quoted uncertainties or 0.5 dex where propagated errors are large.
The parameters used for the XRB sample, and the derived luminosity and power, are listed in Table \ref{table1}.

\begin{table*}
\caption{XRB parameter values used to determine the luminosity, power, and Doppler- and collimation-corrected luminosity. Values in brackets are assumed. The derived luminosity and power values for the sample of XRB. The observed $\gamma$-ray luminosity $L_{\gamma,{\rm obs,iso}}$ is determined from the $\it Fermi$ LAT photon flux and spectral index. The minimum jet power $P_j$, and the Doppler- and collimation-corrected luminosity $L_{\gamma}$ are shown; where two values are present, the first is for the lower Lorentz factor in the parameters, the second for the highest.  References: [1] Pepe, Vila \& Romero 2015. [2] Dubus, Cerutti \& Henri 2010. [3] Tanaka et al. 2016. [4] Miller-Jones, Fender \& Nakar 2006. [5] Bodaghee et al. 2013. [6] Fermi LAT Collaboration 2009. [7] Loh et al. 2016. [8] Reid et al. 2011. [9] Huppenkothen et al. 2017. [10] Fender, Belloni \& Gallo 2004. [11] Orosz et al. 2011. [12] Ling, Zhang \& Tang 2009. [13] Zdziarski et al. 2016. [14] Corbel et al. 2012. [15] Reid et al. 2014}
\label{table1}
\centering
\resizebox{\textwidth}{!}{\begin{tabular}{c | c c c c c}
Parameter & Cyg X-1 & Cyg X-3 & V404 Cyg & GRS1915+105 & GX339-4 \\
\hline
\hline
$\Gamma$ & 1.25$_{[1]}$-3.3$\pm0.1_{[4]}$ & 1.2$_{[2,4]}$-2.0$\pm0.1_{[4]}$ & 2.3$\pm0.5_{[3]}$ & $1.75_{[4]}$ & $4.9_{[4]}$ \\
$N_\gamma$ ($\times 10^{-6}$ ph cm$^{-2}$ s$^{-1}$) & $1.4\pm 0.4_{[5]}$ & $3.5\pm 0.5_{[14]}$ & $2.3\pm 0.8_{[7]}$ & $<0.023\pm0.017_{[5]}$ & $<0.016\pm0.010_{[5]}$ \\
$\alpha$ (ph spec. index) & $2.4_{[13]}$ & $2.7_{[6]}$ & $2.5_{[7]}$ & (2.5) & (2.5)\\ 
$D$ (kpc) & $1.86\pm 0.12_{[8]}$ & $7.0 \pm 0.4_{[12]}$ & $2.39\pm 0.14_{[9]}$ & $9\pm2_{[15]}$ & $8\pm2_{[10]}$ \\
$i$ ($^\circ$) & $27.1\pm0.8_{[11]}$ & $20\pm2_{[2,4]}$ & $67\pm2_{[9]}$ & $60\pm5_{[15]}$ & $80.2\pm0.4_{[4]}$ \\
$S_\nu$ ($\times 10^3$ mJy) & 0.028$_{[10]}$  & 13.4$_{[4]}$ & 3.4$_{[7]}$ & $214_{[10]}$ & $55_{[10]}$ \\
$\nu$ (GHz) & 5.0 & 5.0 & 13.9 & 5.0 & 5.0 \\
$\Delta t$ ($\times 10^4$ s) & 0.2$_{[10]}$ & 30$_{[4]}$ & 6.9$_{[7]}$ & $4.32_{[10]}$ & $1.98_{[10]}$ \\
\hline
$L_{\gamma,{\rm obs,iso}}$ ($\times 10^{35}$ erg s$^{-1}$) & $2.70\pm { 0.34}$ & $76.6 \pm { 8.75}$ & $7.30 \pm { 0.86}$ & $1.00 \pm { 0.43}$ & $0.50 \pm { 0.26}$ \\
$L_{\gamma}$ ($\times 10^{36}$ erg s$^{-1}$) & $0.13 \pm { 0.02}$ & $3.16 \pm { 0.60}$ & $18.8 \pm { 2.20}$ & $0.54 \pm { 0.24}$ & $70.0 \pm { 35.8}$ \\
   & $0.42 \pm { 0.08}$ & $2.52 \pm { 0.64}$ & & & \\
$P_j$ ($\times 10^{37}$ erg s$^{-1}$) & $0.03 \pm 0.01$ & $20.7 \pm 4.89$ & $17.3 \pm12.23 $ & $4.90 \pm 3.42$ & $30.7 \pm 4.20$ \\
   & $0.02 \pm 0.01 $ & $7.70 \pm 1.98$ & & & \\
\hline 
\end{tabular}}
\end{table*}

\section{Discussion}
\label{S5}

Using the observed peak {\it Fermi} LAT $\gamma$-ray flux or upper limit, the jet to line-of-sight inclination, and the jet Lorentz factor, we have made estimates for the on-axis, isotropic equivalent $\gamma$-ray luminosity from the jets of five XRBs.
The isotropic on-axis luminosity is further corrected for the collimated emission, where the fraction is given by $1-\cos(1/\Gamma)$, resulting in a collimation-corrected estimate for the $\gamma$-ray luminosity.
This $\gamma$-ray luminosity, along with an estimate for the jet power, can be directly compared with the universal scaling for relativistic jets from BH systems proposed by \cite{2012Sci...338.1445N}.
The Nemmen relation is based on the peak $\gamma$-ray luminosity and jet power for blazars and $\gamma$-ray bursts (GRB).
The inclusion of XRBs on this plot, extends this $L_\gamma-P_j$ relation to lower luminosities and power.
The XRB fit on this plot can also be used to indicate the jet origin of $\gamma$-ray photons from such sources.

The $\gamma$-ray luminosity for the three source types, blazars, GRBs, and XRBs, is the beamed on-axis and collimation corrected luminosity.
The jet power is estimated for each source type uniquely:
for blazars, the jet power is found from a tight correlation between the radio luminosity and the power required to inflate an X-ray cavity \citep{2010ApJ...720.1066C}.
Using the relation $P_j\sim6\times10^{43}L_{40}^{0.7}$ erg s$^{-1}$, where $L_{40}$ is the radio luminosity in units $\times 10^{40}$ erg s$^{-1}$, the power for blazars with VLA observed extended radio emission was determined;
for GRBs, the jet power is found using a collimation corrected estimate for the kinetic energy from the peak of the radio or X-ray afterglow and assuming the fireball model.
The jet power is then $P_j =(1+z)f_b E_{k,{\rm iso}}/t_{90}$, where $f_b$ is the collimation correction, $E_{k,{\rm iso}}$ the isotropic equivalent kinetic energy, and $t_{90}$ the timescale for 90\% of the prompt emission energy;
for XRBs, the jet power is found using the minimum energy assuming equipartition and the peak radio flare flux density.
The jet power is given by equation \ref{PJ}.  
Our estimates for the Doppler- and collimation-corrected $\gamma$-ray luminosity and jet power, for the five XRBs in our sample, all fall within the uncertainties associated with the original $L_\gamma-P_j$ relation for BH jets: 
$\log{P_j}= (0.98\pm0.02)\log{L_\gamma}+(1.6\pm0.9)~{~\rm erg~s^{-1}}$.

The $L_\gamma-P_j$ correlation can be applied to XRBs without a limit on the $\gamma$-ray luminosity.
There are at least four additional XRBs with peak radio flare flux densities, rise times, $\beta\cos i$, and distance measurements:
GRO J1655-40, V4641 Sgr, XTE J1550-564, and H 1743-322.
All have BH components \citep{2016A&A...587A..61C,2016ApJS..222...15T}.
The distances to these systems are:
GRO J1655-40 is at 3.2 kpc \citep{1995Natur.375..464H};
V4641 Sgr is at 6.2 kpc \citep{2014ApJ...784....2M};
XTE J1550-564 is at 4.5 kpc \citep{2011ApJ...730...75O};
H 1743-322 is at 10 kpc \citep{2009ApJ...699..453S,2009ApJ...698.1398M}.
Given the $\beta\cos i$ and $\mu_a\mu_r$ values in MFN06, the bulk Lorentz factors are $\Gamma=[2.5, ~\geq2.5, ~1.3, ~3.7]$ respectively.
The power of the jet for these systems, using the rise time and peak flux listed in MFN06 for GRO J1655-40, V4641 Sgr, and XTE J1550-564, and the rise time and peak flux from \cite{2009ApJ...698.1398M} for H 1743-322, is then $P_j=[2.93, ~1.05, ~0.04, ~8.33]\times 10^{38}$ erg s$^{-1}$ respectively.
The $L_\gamma-P_j$ relation can give us contraints on the on-axis $\gamma$-ray luminosity.
As the observed upper limit for $\gamma$-ray luminosity is $L_{\gamma,{\rm obs,iso}}=f_b^{-1} a^3 L_\gamma$, using equation \ref{E1}, the maximum $\gamma$-ray photon flux at a detector for each source is:
$N_\gamma\leq [1.6, ~0.5, ~2.8, ~0.1]\times 10^{-8}$ photons s$^{-1}$ cm$^{-2}$ respectively, at energies $> 100$ MeV and assuming $\alpha=2.5$.

The inclination angle used for the relativistic Doppler correction is in all cases assumed to be the angle from a point source on the jet-axis to the line-of-sight.
However, the jets have a finite opening angle $\phi$;
the angle to the jet could be as low as ($i-\phi$).
The Doppler corrected luminosity will be lower in each case than those presented here.
Cyg X-1 and Cyg X-3 have relatively small inclination angles, $27^\circ.1$ and $\sim20^\circ$, and jet opening angles, $<18^\circ$ and $<16^\circ.5$ respectively.
The Doppler and collimation corrected values for each system using ($i-\phi$) are shifted to lower $\gamma$-ray luminosities.
For Cyg X-1, $L_\gamma \sim 2.5\times 10^{34}$ erg s$^{-1}$ when $\Gamma=3.3$, and for both $\Gamma$ values used here is closer to the central $L_\gamma-P_j$ trend.
For Cyg X-3, $L_\gamma \sim 10^{36}$ erg s$^{-1}$ for $\Gamma=2$, and for both $\Gamma$ values used is well within the correlation limits.

However, note that the Doppler-corrected $\gamma$-ray energies for the five XRBs are most likely underestimates;
this is due to the shift of the observed {\it Fermi} LAT band $>100$ MeV, to the on-axis energy range, where $\nu_{\rm obs}=a\nu_{\rm o}$ and $\nu_{\rm o}$ is the value to an on-axis observer.
The observed {\it Fermi} LAT spectrum, in all cases, is assumed to be a single power law;
without information regarding the spectral peak or index below the observed minimum energy 100 MeV, we have assumed the on-axis $\gamma$-ray luminosity to be equivalent to the energy in the Doppler-corrected band i.e. a flat spectrum.
If the single power-law extended to lower energies than those observed by {\it Fermi} LAT then the on-axis Doppler corrected luminosities would be of order $L_\gamma \sim 10^{41}$ erg s$^{-1}$;
such a bright on-axis source could be detectable as a $\gamma$-ray transient in local galaxies e.g. $N_\gamma\sim 2\times 10^{-6}$ ph s$^{-1}$ cm$^{-2}$ at 1 Mpc, and becoming limited at $N_\gamma\sim 2\times 10^{-8}$ ph s$^{-1}$ cm$^{-2}$ at 10 Mpc.

To estimate the minimum power of the jet we have assumed a ratio of energy in relativistic protons to electrons in the synchrotron emitting region of $\epsilon_p/\epsilon_e=1$.
This ratio could in reality be very small or as high as $\sim100$ e.g. GRBs, where the ratio is typically in the range $10\la\epsilon_p/\epsilon_e\la100$.
If the energy in the hadronic particles is larger, the jet powers presented here would be underestimates;
for $\epsilon_p/\epsilon_e=2$ the jet power would increase by a factor of $\sim1.3$, for $\epsilon_p/\epsilon_e=100$ the power would be $\sim9.4$ times those presented here.
Alternatively, if the energy in relativistic protons is very small, then the jet power would be $\sim0.7$ of those presented.
As noted by \cite{2014MNRAS.445.1321Z}, this method does not consider the contribution by cold ions in the jet bulk flow to the total power.
The minimum jet powers presented here are therefore underestimates;
the maximum correction factor to the presented powers is a factor $\sim 50$ larger.
If the minimum jet powers presented here are massively underestimated, then by considering similar arguments for the underestimate of the jet power in blazars \citep[e.g.][]{1999ASPC..161..249G} the $L_\gamma-P_j$ correlation may still hold but with a shallower index.

\begin{figure*}
\includegraphics[scale=0.8]{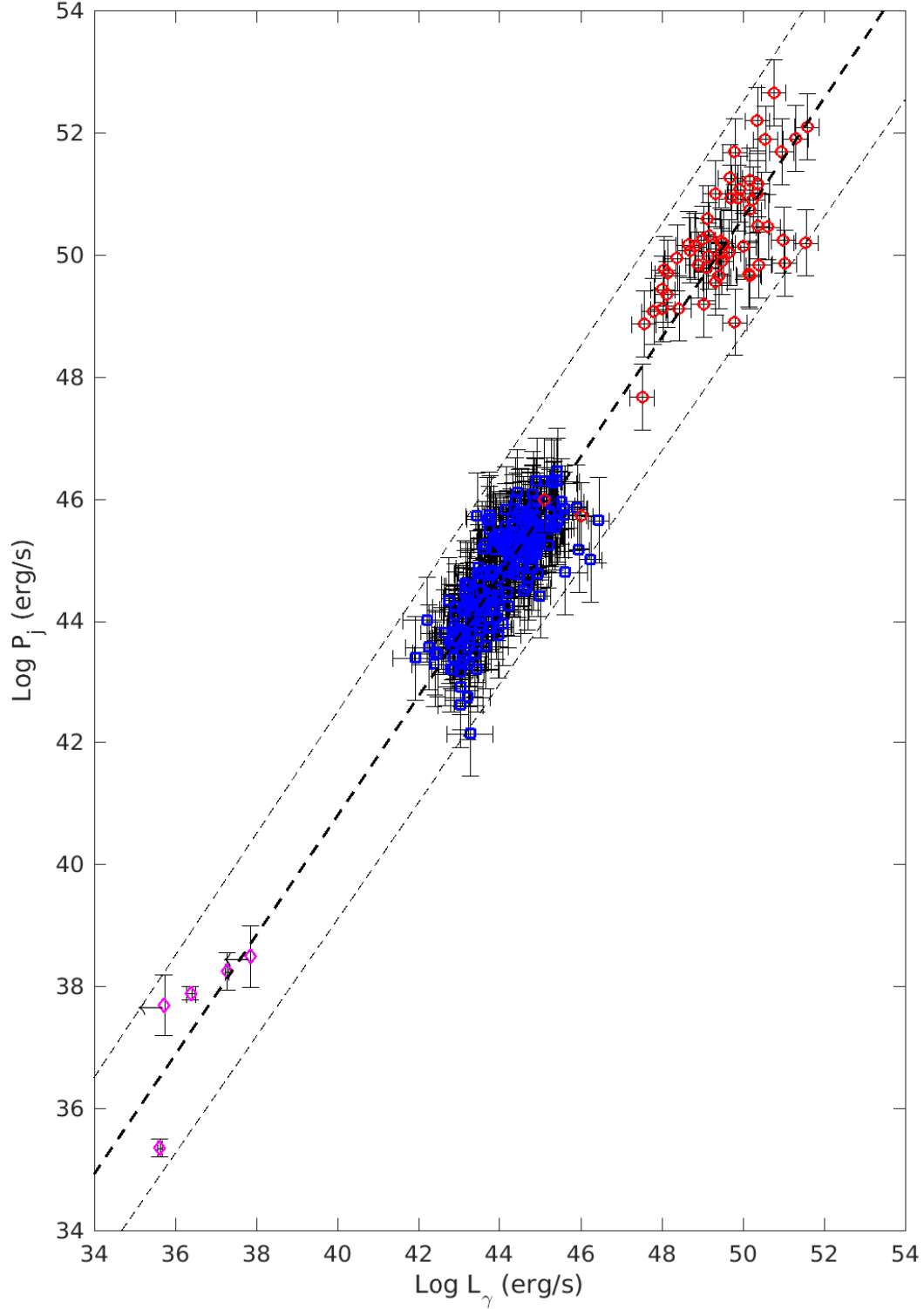}
\centering
\caption{$L_\gamma-P_j$ relation including five XRBs. Doppler- and collimation-corrected luminosity and power estimates are shown for XRB (pink diamonds), using the larger $\Gamma$ value where appropriate. Values for blazars are shown as blue squares, and $\gamma$-ray bursts are shown as red circles, where the data and uncertainties are from Nemmen et al. 2012. Dashed black line is the $L_\gamma-P_j$ relation, thin dashed lines represent the limits from Nemmen et al. 2012.}
\label{fig2}
\end{figure*}

Figure \ref{fig2} shows the \cite{2012Sci...338.1445N} distribution of blazars and GRBs with the uncertainties for each population, plus the five XRBs presented here.
Where two estimates for the Doppler-corrected luminosity and power exist for an XRB, we have used the values that correspond to the largest $\Gamma$.
The addition of more XRBs to this distribution will help to determine the validity of the correlation, and if it holds, better constrain the index and limits for a wide range of BH jets in $L_\gamma-P_j$.

\cite{2014ApJ...780L..14M} found a similar correlation for XRBs in the hard state using the bolometric luminosity for the jet derived from models;
the power estimates for the jets in their sample were typically lower than those found here by up to 3-4 orders of magnitude.
Our estimates are based on the minimum jet power during a flaring/transient event as opposed to the compact jets seen during the hard state;
this difference can explain the disagreement in jet power where the same source is compared.
The luminosity used in our sample is the $\gamma$-ray flare luminosity not the hard-state bolometric jet luminosity and therefore our estimates are directly comparable to the original \cite{2012Sci...338.1445N} correlation.

That \cite{2014ApJ...780L..14M} find a correlation without using $\gamma$-ray luminosity demonstrates that a common mechanism links all BHs and jets through accretion with very small differences.
By considering only the $\gamma$-ray flux from these jets we can probe the part of the outflow with the highest Lorentz factor and strongest relativistic beaming.
For GRBs, the emitted $\gamma$-rays are a small fraction of the total engine energy; 
despite the differences in these sources (stellar mass BH in XRBs, SMBH in AGN, or SNe/merger for GRBs) the observed relation is always the same.
A confirmed correlation for $L_\gamma-P_j$ for jets from accreting BH systems, regardless of phenomenological differences between the systems, could help determine a ubiquitous emission mechanism for high energy photons from such jets.
The existence of a correlation across extremes of time and mass scale points to common physical phenomena between all relativistic BH jets.
If all relativistic BH jets have the same high-energy emission mechanism then the differences between the system classes can be used to constrain the emission mechanism at $\gamma$-ray energies.
Alternatively, the correlation may indicate that the efficiency for various $\gamma$-ray emission processes in relativistic jets is similar.

Two groups of models are used to explain the high energy emission in XRBs and blazars:
the hadronic/lepto-hadronic models, where the high energy emission is from internal jet processes such as synchrotron self Compton (SSC), synchrotron of protons or the decay of neutral pions from proton-proton cascade;
and the leptonic models, where high energy emission is due to the external Compton scattering by relativistic electrons of a strong photon field, either from stellar companion black-body photons or X-ray photons from the accretion disk for XRBs, or the accretion disk and broadline region for blazars.

Strong polarization measurements made in the $\gamma$-ray tail of Cyg X-1 favour a lepto-hadronic model, and a jet origin, for the high energy emission \citep{2015ApJ...807...17R}.
Lepto-hadronic models for the broadband emission of Cyg X-1 by \cite{2015A&A...584A..95P} also favours a synchrotron or SSC, and therefore jet origin, for the high energy tail.
The low mass of the companion to V404 Cyg, the temporal association of $\gamma$-ray excess and radio flares, and the simultaneous detection of the 511 keV annihilation line and $\gamma$-rays of higher energy all point to the jet as the origin for such emission \citep{2016MNRAS.462L.111L}.
Long-term monitoring of blazars indicates a correlation between $\gamma$-ray flares and optical flares; the optical emission from blazars is polarized to different degrees depending on the location of the synchrotron peak relative to the observed optical bands \citep{2016MNRAS.462.4267J}; optical and $\gamma$-ray flares, and high polarizations, are associated with the jet.
Polarization measurements of the early afterglow in GRBs indicates the existence of an ordered magnetic field in a magnetized baryonic jet \citep{2009Natur.462..767S,2013Natur.504..119M}.
Alternatively, if high energy emission from AGN and XRBs is due to a leptonic process i.e. inverse Compton scattering of external photons, then this correlation may have implications for the emission mechanism responsible for the high-energy prompt GRB.
For a long GRB the target photons may be from shock breakout, early supernova photosphere photons, or the photons of a companion star whose presence may be inferred by the high degree of stripping of long GRB progenitor supernovae i.e. Ic SNe.

Throughout, we have made the assumption that all GRBs are powered by BH, as opposed to magnetars;
the overall $L_\gamma-P_j$ correlation presented here may support this assumption.
XRBs, blazars, and GRBs all populating the same relation for $L_\gamma-P_j$ indicates a common jet emission mechanism, or efficiency, for $\gamma$-rays where the magnitude of $L_\gamma$ depends on the jet power.

\section{Conclusions}
\label{S6}

We have shown that when corrected for collimation and Doppler boosting, the $\gamma$-ray luminosity from XRBs follows the $L_\gamma-P_j$ relation found by \cite{2012Sci...338.1445N} for relativistic jets from BH systems.
This correlation holds across $\sim 17$ orders of magnitude and is the first attempt at comparing the energetics, using $\gamma$-ray luminosities, for three classes of accreting BH systems e.g. XRB, AGN, and GRB.
Although the jet powers and $\gamma$-ray luminosities for XRBs are most likely underestimates, XRBs are relatively closely grouped in the parameter space.
The power of a jet from a BH system can be independently constrained by the on-axis $\gamma$-ray luminosity.
Alternatively, the jet power can be used to indicate the expected on-axis $\gamma$-ray luminosity for high energy flares from BH jets.
Future target of opportunity high energy observations of XRB during radio flaring events could help further constrain this relation.
If such a relation is ubiquitous amongst relativistic jets from BHs, then a common emission mechanism, or efficiency, is most likely responsible.
By comparing the different systems, constraints can be put on the emission dynamics.

\section*{Acknowledgements}

GPL thanks Iain Steele and Phil James for helpful comments, and the participants of the Astrophysical Jets School of Carg\'ese 2016 for useful discussions.
This research was supported by STFC grants.
EP acknowledges financial support from INAF-ASI contract I/088/06/0 from the Italian Ministry of Education and Research, and from Scuola Normale Superiore.

\bsp
\label{lastpage}

\begin{thebibliography}{999}
\bibitem[Blandford \& K{\"o}nigl(1979)]{1979ApJ...232...34B} Blandford, R.~D., \& K{\"o}nigl, A.\ 1979, \apj, 232, 34 
\bibitem[Bodaghee et al.(2013)]{2013ApJ...775...98B} Bodaghee, A., Tomsick, J.~A., Pottschmidt, K., et al.\ 2013, \apj, 775, 98 
\bibitem[Burbridge (1956)]{1956Phys.Rev...103...264}Burbridge, G. R.\ 1956, Phys. Rev., 103, 264
\bibitem[Burbridge (1959)]{1959ApJ...129...849}Burbridge, G. R.\ 1959, \apj, 129, 849
\bibitem[Casares \& Charles(1994)]{1994MNRAS.271L...5C} Casares, J., \& Charles, P.~A.\ 1994, \mnras, 271, L5
\bibitem[Cavagnolo et al.(2010)]{2010ApJ...720.1066C} Cavagnolo, K.~W., McNamara, B.~R., Nulsen, P.~E.~J., et al.\ 2010, \apj, 720, 1066
\bibitem[Corbel et al.(2000)]{2000A&A...359..251C} Corbel, S., Fender, R.~P., Tzioumis, A.~K., et al.\ 2000, \aap, 359, 251
\bibitem[Corbel et al.(2003)]{2003A&A...400.1007C} Corbel, S., Nowak, M.~A., Fender, R.~P., Tzioumis, A.~K., \& Markoff, S.\ 2003, \aap, 400, 1007
\bibitem[Corbel et al.(2012)]{2012MNRAS.421.2947C} Corbel, S., Dubus, G., Tomsick, J.~A., et al.\ 2012, \mnras, 421, 2947 
\bibitem[Corbel et al.(2013)]{2013MNRAS.428.2500C} Corbel, S., Coriat, M., Brocksopp, C., et al.\ 2013, \mnras, 428, 2500
\bibitem[Corral-Santana et al.(2016)]{2016A&A...587A..61C} Corral-Santana, J.~M., Casares, J., Mu{\~n}oz-Darias, T., et al.\ 2016, \aap, 587, A61
\bibitem[Dubus et al.(2010)]{2010MNRAS.404L..55D} Dubus, G., Cerutti, B., \& Henri, G.\ 2010, \mnras, 404, L55
\bibitem[Falcke et al.(2004)]{2004A&A...414..895F} Falcke, H., K{\"o}rding, E., \& Markoff, S.\ 2004, \aap, 414, 895
\bibitem[Fender \& Pooley(1998)]{1998MNRAS.300..573F} Fender, R.~P., \& Pooley, G.~G.\ 1998, \mnras, 300, 573
\bibitem[Fender(2001)]{2001MNRAS.322...31F} Fender, R.~P.\ 2001, \mnras, 322, 31
\bibitem[Fender \& Belloni(2004)]{2004ARA&A..42..317F} Fender, R., \& Belloni, T.\ 2004, \araa, 42, 317
\bibitem[Fender et al.(2004)]{2004MNRAS.355.1105F} Fender, R.~P., Belloni, T.~M., \& Gallo, E.\ 2004, \mnras, 355, 1105
\bibitem[Fermi LAT Collaboration et al.(2009)]{2009Sci...326.1512F} Fermi LAT Collaboration, Abdo, A.~A., Ackermann, M., et al.\ 2009, Science, 326, 1512
\bibitem[Ghisellini(1999)]{1999ASPC..161..249G} Ghisellini, G.\ 1999, High Energy Processes in Accreting Black Holes, 161, 249
\bibitem[Granot et al.(2002)]{2002ApJ...570L..61G} Granot, J., Panaitescu, A., Kumar, P., \& Woosley, S.~E.\ 2002, \apjl, 570, L61
\bibitem[Heinz \& Sunyaev(2003)]{2003MNRAS.343L..59H} Heinz, S., \& Sunyaev, R.~A.\ 2003, \mnras, 343, L59
\bibitem[Hjellming \& Rupen(1995)]{1995Natur.375..464H} Hjellming, R.~M., \& Rupen, M.~P.\ 1995, \nat, 375, 464
\bibitem[Huppenkothen et al.(2017)]{2017ApJ...834...90H} Huppenkothen, D., Younes, G., Ingram, A., et al.\ 2017, \apj, 834, 90
\bibitem[Jermak et al.(2016)]{2016MNRAS.462.4267J} Jermak, H., Steele, I.~A., Lindfors, E., et al.\ 2016, \mnras, 462, 4267
\bibitem[Jorstad et al.(2013)]{2013ApJ...773..147J} Jorstad, S.~G., Marscher, A.~P., Smith, P.~S., et al.\ 2013, \apj, 773, 147
\bibitem[Lewin \& van der Klis(2006)]{2006csxs.book.....L} Lewin, W.~H.~G., \& van der Klis, M.\ 2006, Compact stellar X-ray sources, 39
\bibitem[Ling et al.(2009)]{2009ApJ...695.1111L} Ling, Z., Zhang, S.~N., \& Tang, S.\ 2009, \apj, 695, 1111
\bibitem[Lister et al.(2009)]{2009AJ....138.1874L} Lister, M.~L., Cohen, M.~H., Homan, D.~C., et al.\ 2009, \aj, 138, 1874-1892
\bibitem[Longair(1994)]{1994hea2.book.....L} Longair, M.~S.\ 1994, High energy astrophysics.~Volume 2.~Stars, the Galaxy and the interstellar medium., by Longair, M.~S..~ Cambridge University Press, Cambridge (UK), 1994, 410 p., ISBN 0-521-43439-4.
\bibitem[Loh et al.(2016)]{2016MNRAS.462L.111L} Loh, A., Corbel, S., Dubus, G., et al.\ 2016, \mnras, 462, L111
\bibitem[Ma et al.(2014)]{2014ApJ...780L..14M} Ma, R., Xie, F.-G., \& Hou, S.\ 2014, \apjl, 780, L14 
\bibitem[MacDonald et al.(2014)]{2014ApJ...784....2M} MacDonald, R.~K.~D., Bailyn, C.~D., Buxton, M., et al.\ 2014, \apj, 784, 2
\bibitem[Marscher et al.(2002)]{2002Natur.417..625M} Marscher, A.~P., Jorstad, S.~G., G{\'o}mez, J.-L., et al.\ 2002, \nat, 417, 625
\bibitem[Marscher(2006)]{2006AIPC..856....1M} Marscher, A.~P.\ 2006, Relativistic Jets: The Common Physics of AGN, Microquasars, and Gamma-Ray Bursts, 856, 1
\bibitem[McClintock et al.(2009)]{2009ApJ...698.1398M} McClintock, J.~E., Remillard, R.~A., Rupen, M.~P., et al.\ 2009, \apj, 698, 1398
\bibitem[Merloni et al.(2003)]{2003MNRAS.345.1057M} Merloni, A., Heinz, S., \& di Matteo, T.\ 2003, \mnras, 345, 1057
\bibitem[M{\'e}sz{\'a}ros(2002)]{2002ARA&A..40..137M} M{\'e}sz{\'a}ros, P.\ 2002, \araa, 40, 137
\bibitem[Miller-Jones et al.(2006)]{2006MNRAS.367.1432M} Miller-Jones, J.~C.~A., Fender, R.~P., \& Nakar, E.\ 2006, \mnras, 367, 1432
\bibitem[Mirabel et al.(1998)]{1998A&A...330L...9M} Mirabel, I.~F., Dhawan, V., Chaty, S., et al.\ 1998, \aap, 330, L9
\bibitem[Mirabel \& Rodr{\'{\i}}guez(1999)]{1999ARA&A..37..409M} Mirabel, I.~F., \& Rodr{\'{\i}}guez, L.~F.\ 1999, \araa, 37, 409
\bibitem[Mundell et al.(2013)]{2013Natur.504..119M} Mundell, C.~G., Kopa{\v c}, D., Arnold, D.~M., et al.\ 2013, \nat, 504, 119
\bibitem[Mu{\~n}oz-Darias et al.(2008)]{2008MNRAS.385.2205M} Mu{\~n}oz-Darias, T., Casares, J., \& Mart{\'{\i}}nez-Pais, I.~G.\ 2008, \mnras, 385, 2205
\bibitem[Nemmen et al.(2012)]{2012Sci...338.1445N} Nemmen, R.~S., Georganopoulos, M., Guiriec, S., et al.\ 2012, Science, 338, 1445
\bibitem[Orosz et al.(2011)]{2011ApJ...730...75O} Orosz, J.~A., Steiner, J.~F., McClintock, J.~E., et al.\ 2011, \apj, 730, 75
\bibitem[Orosz et al.(2011)]{2011ApJ...742...84O} Orosz, J.~A., McClintock, J.~E., Aufdenberg, J.~P., et al.\ 2011, \apj, 742, 84
\bibitem[Pepe et al.(2015)]{2015A&A...584A..95P} Pepe, C., Vila, G.~S., \& Romero, G.~E.\ 2015, \aap, 584, A95
\bibitem[Piano et al.(2017)]{2017arXiv170310085P} Piano, G., Munar-Adrover, P., Verrecchia, F., Tavani, M., \& Trushkin, S.~A.\ 2017, arXiv:1703.10085 
\bibitem[Piran(2004)]{2004RvMP...76.1143P} Piran, T.\ 2004, Reviews of Modern Physics, 76, 1143
\bibitem[Reid et al.(2011)]{2011ApJ...742...83R} Reid, M.~J., McClintock, J.~E., Narayan, R., et al.\ 2011, \apj, 742, 83
\bibitem[Reid et al.(2014)]{2014ApJ...796....2R} Reid, M.~J., McClintock, J.~E., Steiner, J.~F., et al.\ 2014, \apj, 796, 2
\bibitem[Rodriguez et al.(2003)]{2003ApJ...595.1032R} Rodriguez, J., Corbel, S., \& Tomsick, J.~A.\ 2003, \apj, 595, 1032
\bibitem[Rodriguez et al.(2008)]{2008ApJ...675.1449R} Rodriguez, J., Shaw, S.~E., Hannikainen, D.~C., et al.\ 2008, \apj, 675, 1449-1458
\bibitem[Rodriguez et al.(2015)]{2015ApJ...807...17R} Rodriguez, J., Grinberg, V., Laurent, P., et al.\ 2015, \apj, 807, 17
\bibitem[Rybicki \& Lightman(1986)]{1986rpa..book.....R} Rybicki, G.~B., \& Lightman, A.~P.\ 1986, Radiative Processes in Astrophysics, by George B.~Rybicki, Alan P.~Lightman, pp.~400.~ISBN 0-471-82759-2.~Wiley-VCH , June 1986., 400
\bibitem[Saikia et al.(2016)]{2016MNRAS.461..297S} Saikia, P., K{\"o}rding, E., \& Falcke, H.\ 2016, \mnras, 461, 297
\bibitem[Sari et al.(1999)]{1999ApJ...519L..17S} Sari, R., Piran, T., \& Halpern, J.~P.\ 1999, \apjl, 519, L17
\bibitem[Shaposhnikov \& Titarchuk(2009)]{2009ApJ...699..453S} Shaposhnikov, N., \& Titarchuk, L.\ 2009, \apj, 699, 453
\bibitem[Sironi \& Spitkovsky(2011)]{2011ApJ...726...75S} Sironi, L., \& Spitkovsky, A.\ 2011, \apj, 726, 75
\bibitem[Steele et al.(2009)]{2009Natur.462..767S} Steele, I.~A., Mundell, C.~G., Smith, R.~J., Kobayashi, S., \& Guidorzi, C.\ 2009, \nat, 462, 767
\bibitem[Szostek et al.(2008)]{2008MNRAS.388.1001S} Szostek, A., Zdziarski, A.~A., \& McCollough, M.~L.\ 2008, \mnras, 388, 1001
\bibitem[Tanaka et al.(2016)]{2016ApJ...823...35T} Tanaka, Y.~T., Itoh, R., Uemura, M., et al.\ 2016, \apj, 823, 35
\bibitem[Tavani et al.(2009)]{2009Natur.462..620T} Tavani, M., Bulgarelli, A., Piano, G., et al.\ 2009, \nat, 462, 620 
\bibitem[Tetarenko et al.(2016)]{2016ApJS..222...15T} Tetarenko, B.~E., Sivakoff, G.~R., Heinke, C.~O., \& Gladstone, J.~C.\ 2016, \apjs, 222, 15
\bibitem[T{\"u}rler et al.(2004)]{2004A&A...415L..35T} T{\"u}rler, M., Courvoisier, T.~J.-L., Chaty, S., \& Fuchs, Y.\ 2004, \aap, 415, L35 
\bibitem[Urry \& Padovani(1995)]{1995PASP..107..803U} Urry, C.~M., \& Padovani, P.\ 1995, \pasp, 107, 803
\bibitem[Vilhu \& Hannikainen(2013)]{2013A&A...550A..48V} Vilhu, O., \& Hannikainen, D.~C.\ 2013, \aap, 550, A48
\bibitem[Wang \& Dai(2017)]{2017MNRAS.470.1101W} Wang, F.~Y., \& Dai, Z.~G.\ 2017, \mnras, 470, 1101
\bibitem[Yuan \& Zhang(2012)]{2012ApJ...757...56Y} Yuan, F., \& Zhang, B.\ 2012, \apj, 757, 56
\bibitem[Zanin et al.(2016)]{2016A&A...596A..55Z} Zanin, R., Fern{\'a}ndez-Barral, A., de O{\~n}a Wilhelmi, E., et al.\ 2016, \aap, 596, A55
\bibitem[Zdziarski et al.(2014)]{2014MNRAS.442.3243Z} Zdziarski, A.~A., Pjanka, P., Sikora, M., \& Stawarz, {\L}.\ 2014, \mnras, 442, 3243
\bibitem[Zdziarski(2014)]{2014MNRAS.445.1321Z} Zdziarski, A.~A.\ 2014, \mnras, 445, 1321
\bibitem[Zdziarski et al.(2016)]{2016arXiv160705059Z} Zdziarski, A.~A., Malyshev, D., Chernyakova, M., \& Pooley, G.~G.\ 2016, arXiv:1607.05059
\end{thebibliography}
\end{document}